\documentclass[aps,superscriptaddress,twocolumn,showpacs]{revtex4-1}
\pdfoutput=1
\usepackage{color}
\usepackage{amsmath, amsthm, amsfonts}    % need for subequations
\usepackage{amssymb}
\usepackage{mathtools}
\usepackage{mathptmx} 
\usepackage[table]{xcolor}
\usepackage{dsfont}
\usepackage{enumitem} 
\usepackage{appendix}
\usepackage{braket}
\usepackage{cancel}
\usepackage{lipsum}% http://ctan.org/pkg/lipsum
\usepackage{graphicx}% http://ctan.org/pkg/graphicx
\usepackage{tikz}
\usepackage{makecell}
\usepackage[unicode=true,pdfusetitle,
 bookmarks=true,bookmarksnumbered=false,bookmarksopen=false,
 breaklinks=true,pdfborder={0 0 0},backref=false,colorlinks=true,citecolor=blue]{hyperref}
\usepackage{graphicx}   % need for figures
\usepackage[caption=false]{subfig}       % use for side-by-side figures and   to have captions justified and small letter size
\usepackage{bm}            % bold math
\usepackage[normalem]{ulem}  % use for underlining

\newlength\imagewidth
\newlength\imagescale
\def\be{\begin{eqnarray}}
\def\ee{\end{eqnarray}}
\def\r{{\bf r}}

\def\E{{\bf E}}
\def\H{{\bf H}}

\def\im{{\rm i}}

\DeclareUnicodeCharacter{2212}{-}

\definecolor{JOT-color}{named}{blue}
\definecolor{CSF-color}{named}{orange}

\begin{document}
 
\title{Capturing near-field circular dichroism  enhancements from far-field measurements}

%\title{The role of optical losses in near-field circular dichroism enhancements}

%\title{Genetic algorithms towards finding the optimal building block for enhanced chiral detection.}

%\title{Chiral Anapoles}

\author{Jorge Olmos-Trigo}
\email{jolmostrigo@gmail.com}
\affiliation{Donostia International Physics Center, Paseo Manuel de Lardizabal 4, 20018 Donostia-San Sebastian, Spain.}
\affiliation{Centro de F\'isica de Materiales, Paseo Manuel de Lardizabal 5, 20018 Donostia-San Sebastian, Spain.}

\author{Jon Lasa-Alonso}
\affiliation{Donostia International Physics Center, Paseo Manuel de Lardizabal 4, 20018 Donostia-San Sebastian, Spain.}
\affiliation{Centro de F\'isica de Materiales, Paseo Manuel de Lardizabal 5, 20018 Donostia-San Sebastian, Spain.}

\author{Iker G\'omez-Viloria}
\affiliation{Centro de F\'isica de Materiales, Paseo Manuel de Lardizabal 5, 20018 Donostia-San Sebastian, Spain.}

\author{Gabriel Molina-Terriza}
\affiliation{Donostia International Physics Center, Paseo Manuel de Lardizabal 4, 20018 Donostia-San Sebastian, Spain.}
\affiliation{Centro de F\'isica de Materiales, Paseo Manuel de Lardizabal 5, 20018 Donostia-San Sebastian, Spain.}
\affiliation{IKERBASQUE, Basque Foundation for Science, Mar\'ia D\'iaz de Haro 3, 48013 Bilbao, Spain.}

\author{Aitzol Garc\'ia-Etxarri}
\affiliation{Donostia International Physics Center, Paseo Manuel de Lardizabal 4, 20018 Donostia-San Sebastian, Spain.}
\affiliation{IKERBASQUE, Basque Foundation for Science, Mar\'ia D\'iaz de Haro 3, 48013 Bilbao, Spain.}

%\begin{abstract}
%Molecular Circular dichroism (CD) spectroscopy faces significant limitations due to the inherent weakness of chiroptical light-matter interactions. In this view, resonant optical  antennas constitute a promising solution to this problem since they can be tuned to enhance electromagnetic fields while preserving helicity. However, an exact expression of the CD enhancement factor, $\mathit{f}_{\rm{CD}}$, which is a magnitude that describes the electromagnetic near-field enhancement of scatterers associated with a given helicity, is lacking. 
%Here, we derive an exact multipolar expansion of $\mathit{f}_{\rm{CD}}$, which is valid for an arbitrary scatterer under an arbitrary incident illumination. Particularly, we present analytical expressions which remain valid beyond the usual scenario where dipolar particles are excited by an electromagnetic plane wave. Moreover, we show that regardless of the incident illumination, $\mathit{f}_{\rm{CD}}$  can be related to Stokes parameters that can be measured in far-field, i.e., scattering cross-sections and the helicity expectation value. In addition, we show that in the case of lossless cylindrically symmetric samples, the near-field characterization of $\mathit{f}_{\rm{CD}}$ can be done only from two far-field measurements. Our contribution paves the way for the characterization of devices capable of enhancing molecular CD from far-field measurements.   
%\end{abstract}

\begin{abstract}
Molecular Circular dichroism (CD) spectroscopy faces significant limitations due to the inherent weakness of chiroptical light-matter interactions. In this view, resonant optical  antennas constitute a promising solution to this problem since they can be tuned to increase the CD enhancement factor, $\mathit{f}_{\rm{CD}}$, a magnitude describing the electromagnetic near-field enhancement of scatterers associated with a given helicity. 
Here, we derive an exact multipolar expansion of $\mathit{f}_{\rm{CD}}$, which is valid to deduce the integrated near-field CD enhancements of chiral molecules in the presence of scatterers of any size and shape under general illumination conditions.  Based on our analytical findings, we show that the near-field $\mathit{f}_{\rm{CD}}$ factor  can be related to magnitudes that can be computed in the far-field, i.e., the scattering cross-section and the helicity expectation value. Moreover, we show that in the case of lossless cylindrically symmetric samples, the near-field $\mathit{f}_{\rm{CD}}$ factor can be inferred experimentally only from two far-field measurements at specific scattering angles. Our contribution paves the way for the experimental characterization of devices capable of enhancing molecular CD spectroscopy.   
\end{abstract}

\maketitle

Chirality is a geometrical property of objects which are not superimposable with its mirror image. Chiral objects are ubiquitous in nature. Many organic molecules, such as glucose and most biological amino acids, are chiral. Not to mention the DNA double helix, which, in its standard form, always twists like a right-handed screw~\cite{watson2012double}.

In the pharmaceutical industry, chiral speciﬁcity is critical because opposite enantiomers, i.e., mirror pairs of chiral molecules, can have beneficial or detrimental biological effects on our organism depending on their handedness. Perhaps the most representative case of this is the so-called ``Thalidomide tragedy" which took place during the late 50s~\cite{mcbride1961thalidomide}. Thalidomide was extensively prescribed to pregnant women due to its benefits in alleviating nausea. Instead, this drug prevented the proper growth of the fetus~\cite{lenz1962thalidomide}. The outcome was that thousands of infants were born with severe congenital malformations. That fatality occurred because Thalidomide is a chiral molecule that was marketed as a 50/50 mixture of R and S enantiomers. While the S enantiomer is effective in alleviating nausea, the R enantiomer is toxic. Thus, drugs consisting of chiral molecules may indeed have  different therapeutic or toxicological effects.

Even if they share the same atomic composition, enantiomer
pairs are indistinguishable when measuring their scalar molecular properties. Their chiral nature is revealed only when interacting with other chiral entities. 
In electromagnetism, the most common chiral observable is helicity~\cite{calkin1965invariance}. 
Chiral molecules show a preferential absorption for fields of opposite helicities (left or right-handed polarized waves with helicity eigenvalues of $\sigma = \pm1$, respectively). In a conventional  Circular dichroism (CD) spectroscopy setup, a chiral molecular solution is sequentially illuminated by fields of opposite helicities, and the total transmitted power is recorded for each case. The CD signal is then computed by taking the difference between these two power measurements in transmission. However,  the inherent weakness of chiroptical responses strongly limits the sensitivity of CD spectroscopy. 

Optical antennas, designed to control the properties of light~\cite{yu2014flat}, are promising candidates to enhance the spectroscopic signals of chiral samples. The underlying phenomenon is that optical antennas can be engineered to enhance electromagnetic fields while preserving the electromagnetic helicity~\cite{wen2016metasurface, abendroth2020helicity, zhang2020helicity}. Many works have explored optical antennas~\cite{zambrana2016tailoring, garcia2013surface, CzajkowskiAntosiewicz+2022+4287+4297} and metasurfaces, namely, flat planar arrays of optical antennas, made of metallic~\cite{fan2010plasmonic, garcia2018enantiomer, poulikakos2018chiral} or/and high refractive index materials~\cite{garcia2011strong, olmos2020optimal} for enhanced chiral sensing~\cite{tseng2020dielectric, warning2021nanophotonic}. Examples of such works can be found, for instance,  in the ultraviolet~\cite{hu2019high}, visible~\cite{hendry2010ultrasensitive, kelly2018controlling,zhang2017amplification, yao2018enhancing,hanifeh2020helicity, a2020chiral, mohammadi2021dual} or near -and far-infrared spectral range~\cite{hendry2012chiral,  solomon2018enantiospecific,graf2019achiral, garcia2019enhanced, lasa2020surface, droulias2020absolute, rui2022surface}. Moreover and quite recently, optical cavities have been also proposed as efficient platforms for enhanced chiral sensing~\cite{feis2020helicity, khanbekyan2022enantiomer, PhysRevApplied.18.044007}. %Essentially, optical cavities are not limited by the optical theorem~\cite{bohren2008absorption}, and hence, optical cavities can enhance  electromagnetic fields with a desired helicity well beyond the limit of metasurfaces~\cite{feis2020helicity}.
%Most of the previous works are based on the physical picture behind  the first Kerker condition~\cite{kerker1983electromagnetic}, in which the electric and magnetic dipolar responses are identical~\cite{nieto2011angle,olmos2020kerker}. At the first Kerker condition, the total transmission of light and the optical helicity conservation are simultaneously reached~\cite{fernandez2013electromagnetic}. Hence, CD spectra can be more efficiently enhanced. 

However, researchers rely exclusively on numerical methods to design enhanced chiral sensing devices, as capturing the vector character of the near-field contribution can be challenging.
Here, we derive an exact multipolar expansion of $\mathit{f}_{\rm{CD}}$, which is valid to deduce the integrated near-field CD enhancements of chiral molecules in the presence of scatterers of any size and shape under general illumination conditions. From our analytical results, 
%In other words, we present an analytical expression of the $\mathit{f}_{\rm{CD}}$ factor which goes beyond the usual picture of a plane-wave impinging on dipolar optical antennas~\cite{hu2019high, hendry2010ultrasensitive, kelly2018controlling,zhang2017amplification,
%yao2018enhancing,garcia2018enantiomer,
%hanifeh2020helicity, a2020chiral, mohammadi2021dual, hendry2012chiral, solomon2018enantiospecific,graf2019achiral, garcia2019enhanced, lasa2020surface, droulias2020absolute,feis2020helicity, khanbekyan2022enantiomer,rui2022surface}. 
we show that $\mathit{f}_{\rm{CD}}$ is proportional to the scattering cross-section and the helicity expectation value, which are experimentally measurable magnitudes in the far-field. In addition to this, we also show that for lossless and cylindrically symmetric scatterers, it is possible to infer the $\mathit{f}_{\rm{CD}}$ factor only with two far-field measurements: the extinction cross-section and the helicity density at specific scattering angles.
%Our findings open the path for experimental verification and characterization of building blocks for CD enhancements from far-field measurements.

To describe the excitation of chiral molecules, we adopt the formalism introduced by Tang and Cohen~\cite{tang2010optical}. The CD signal of a chiral molecule under the illumination of a well-defined helicity field  ($\sigma = \pm 1$) can be computed in vacuum as
\begin{equation} \label{CD_inc}
\rm{CD}_{\rm{inc}}(\r) = - \frac{4}{\epsilon} \text{Im} \{G \} C^\sigma_{\rm{inc}}(\r),
\end{equation}
where $\rm{G}$ is the chiral polarizability of the molecule and  $\rm{C}^{\sigma}_{\rm{inc}}(\r)$ is the incident local density of optical chirality~\cite{poulikakos2019optical},
\be \label{C_inc}
\rm{C}^{\sigma}_{\rm{inc}}(\r) = \frac{k \epsilon}{2} \text{Im} \{\E^{\sigma}_{\rm{inc}}(\r) \cdot Z {\H_{\rm{inc}}^{\sigma}}^* ( \r ) \}.
\ee
Here, $k$ is the radiation wavevector, $\epsilon$ is the electric permittivity of the medium and $Z = \sqrt{\mu /\epsilon}$ its electromagnetic impedance. Moreover, $\E^{\sigma}_{\rm{inc}}(\r)$ and $ {\H_{\rm{inc}}^{\sigma}} ( \r )$ refer to  incident electromagnetic fields with well-defined helicity~\footnote{Notice that under the widely used illumination of circularly-polarized plane-waves, we get $|C^{\sigma}_{\rm{inc}}(\r)| =  k\epsilon  |E_0|^2 / 2¨$, $E_0$ denoting the  amplitude of the electric field~\cite{tang2010optical}. That is, we obtain a scalar that does not depend on the spatial coordinates.} (see Appendix~\ref{incident_Fields} for more details).

\begin{figure}[t!]
    \centering
    \includegraphics[width=\columnwidth]{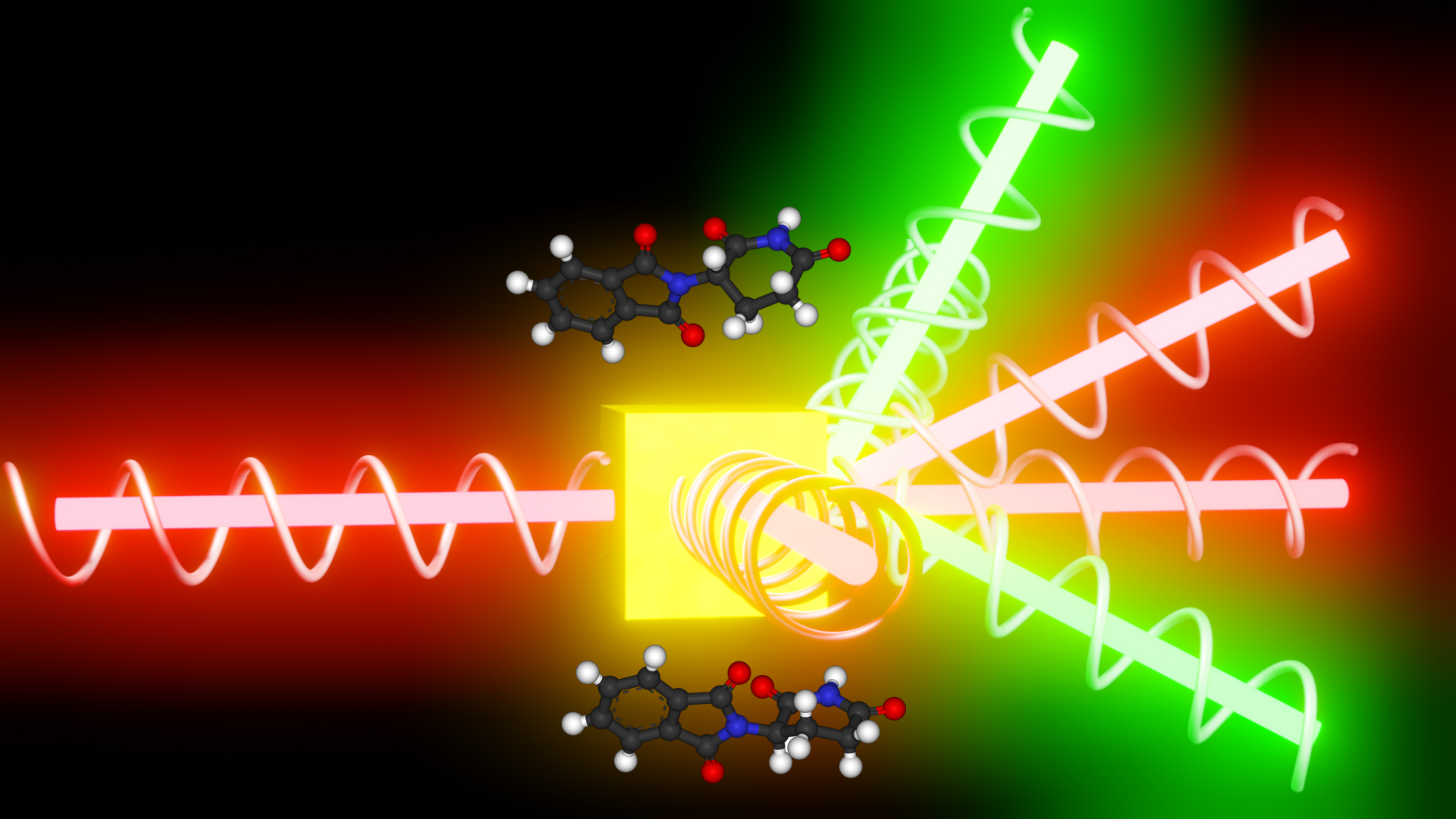}
    \caption{Scattering process in which an incident field with well-defined helicity (red beam with $\sigma = +1$) impinges on an achiral antenna, represented by a glossy cube. Both $\rm{R}$-and-$\rm{S}$ Thalomide enantiomers are also depicted close to the achiral antenna.}
    \label{fig_1}
\end{figure}

In the presence of achiral antennas (see Fig.~\ref{fig_1}) and in the helicity basis~\cite{olmos2019sectoral}, the total electromagnetic fields can be generally written as  $\E^{\sigma \sigma' }_{\rm{tot}}(\r) = \E^{\sigma \sigma' }_{\rm{sca}}(\r) +  \E^{\sigma  }_{\rm{inc}}(\r) \delta_{\sigma \sigma'} $. Here $\E^{\sigma \sigma' }_{\rm{sca}}(\r)$ is the scattered electromagnetic field written in terms of the electromagnetic modes with well defined  helicity $\sigma' = \pm 1$ (see Appendix~\ref{Scattered_Fields} for more details). In analogy with Eq.~\eqref{CD_inc}, we can express the total CD signal of a chiral molecule in the presence of an achiral optical antenna as~\cite{ fernandez2016objects}
\begin{equation} \label{CD_tot}
\rm{CD}_{\rm{tot}}(\r) = - \frac{4}{\epsilon} \text{Im} \{G \} C^\sigma_{\rm{tot}}(\r) =\frac{k}{2} \text{Im} \{G \} \sum_{\sigma' = \pm 1} \sigma' |\E^{\sigma \sigma' }_{\rm{tot}}(\r)|^2.
\end{equation}
We seek to maximize $\rm{CD}_{\rm{tot}}(\r)$ with  Eq.~\eqref{CD_tot}. 
However, such enhancements cannot be achieved through $\rm{G}$ since it is a ﬁxed
chiral molecular parameter that cannot be engineered upon illumination. In contrast, the total density of optical chirality, $\rm{C}^\sigma_{\rm{tot}}(\r)$, can be  tuned to enhance light-matter interactions and thus, increase the sensitivity of CD spectroscopy~\footnote{For the sake of simplicity and hereinafter, we assume the factor $(k \epsilon)/2$ in the definition of the density of optical chirality.}. To get a deeper insight into the total density of optical chirality, let us split $\rm{C}^{\sigma}_{\rm{tot}}(\r)$ into three contributions; namely, $\rm{C}^{\sigma}_{\rm{tot}}(\r)  = \rm{C}^{\sigma}_{\rm{inc}}(\r)+ \rm{C}^{\sigma}_{\rm{sca}}(\r)+\rm{C}^{\sigma}_{\rm{int}}(\r)$, with
\be \label{C_split_inc}
\rm{C}^{\sigma}_{\rm{inc}}(\r)  &=& \sigma |\E^{\sigma}_{\rm{inc}}(\r)|^2 ,\\ \label{C_split_sca} \rm{C}^{\sigma}_{\rm{sca}}(\r)  &=&  |\E^{\sigma + }_{\rm{sca}}(\r)|^2 - |\E^{\sigma - }_{\rm{sca}}(\r)|^2, \\
\label{C_split_ext}
\rm{C}^{\sigma}_{\rm{int}}(\r)  &=& 2 \sigma  \text{Re} \{{\E^{\sigma}_{\rm{inc}}}^* ( \r ) \cdot {\E^{\sigma \sigma}_{\rm{sca}}} ( \r ) \}.
\ee 
%Equations~\eqref{CD_tot}-\eqref{C_split_ext} confirm the suitability
%of the helicity basis to describe the total density of optical chirality. In fact, we can notice that in order to \emph{ideally} enhance $\rm{C}^\sigma_{\rm{tot}}(\r)$, the achiral antenna \emph{must} simultaneously resonate while preserving the incident helicity. From now on, we will refer to these special achiral antennas as resonant helicity-preserving objects. These objects satisfy $|\E^{\sigma \sigma' }_{\rm{sca}}(\r)| = 0$ for $\sigma \neq \sigma'$ whereas $|\E^{\sigma \sigma' }_{\rm{sca}}(\r)|/E_0 \gg 1$ for $\sigma = \sigma'$. The latter gives us insight into that the key to enhancing the total CD signal of a chiral molecule  lies in the full control of the  electromagnetic fields scattered by the achiral antenna, which is governed by Eq.~\eqref{C_split_sca}. Notice that the interference term, given by  Eq.~\eqref{C_split_ext}, contains  incident and scattered fields of the \textit{same} helicity. 

%Hitherto, we have explored requirements to locally enhance CD to conclude that the best candidates are resonant helicity-preserving objects. However, 
It is experimentally challenging to place enantiomers at will, namely,  at a desired spatial coordinate $\r$. Accordingly, it is more convenient to introduce an averaged expression of the local CD enhancement factor that gives insight into how efficient the nanoantenna might be at enhancing CD spectroscopy~\cite{lasa2020surface}. To that end, let us first integrate both Eq.~\eqref{CD_inc} and Eq.~\eqref{CD_tot} over an imaginary sphere of radius $r$ surrounding the achiral antenna to calculate then the ratio between these integrals, namely, 
\begin{equation} \label{f_CD_int}
\mathit{f}_{\rm{CD}} = \frac{ \int {\rm{CD}}_{\rm{tot}}(\r) d S}{\int {\rm{CD}}_{\rm{inc}}(\r) d S}  = 1 + \frac{ \int {\rm{C}}^\sigma_{\rm{sca}}(\r) d S + \int {\rm{C}}^\sigma_{\rm{int}}(\r) d S}{\int {\rm{C}}^\sigma_{\rm{inc}}(\r) d S} ,
\end{equation}
where $dS = r^2 \sin \theta d \varphi d \theta$. 
To compute Eq.~\eqref{f_CD_int}, we need the orthogonality relations that the incident and scattered electromagnetic fields satisfy when written in terms of the multipoles with well-defined helicity. Fortunately, these can be derived from Jackson's book in its third edition~\cite{jackson1999electrodynamics} (see Appendix~\ref{Orthogonality} for the explicit derivation). Now, by considering these relations, and after some algebra (see Appendix~\ref{EXACT} for more details),   we arrive at
\begin{equation} \label{f_CD_int_sep}
\mathit{f}_{\rm{CD}} = 1 + \frac{ \tilde{\rm{C}}^{\sigma}_{\rm{sca}} + \tilde{\rm{C}}^{\sigma}_{\rm{int}}}{\tilde{\rm{C}}^{\sigma}_{\rm{inc}}},
\end{equation}
where 
\be \label{C_incident}
\tilde{\rm{C}}^{\sigma}_{\rm{inc}}  &=& \sigma  \sum_{lm} 
|C^{\sigma }_{lm}|^2 G_{j_l j_l}, \\
\label{C_scattering}
\tilde{\rm{C}}^{\sigma}_{\rm{sca}}  &=&  \sigma \sum_{lm} |C^{\sigma }_{lm}|^2  \text{Re} \{a_{lm} {b_{lm}}^* \} G_{h_l h_l},\\ \label{C_interaccion}
\tilde{\rm{C}}^{\sigma}_{\rm{int}} &=&   \sigma \sum_{lm} |C^{\sigma }_{lm}|^2  \text{Re} \{  \left( a_{lm} + b_{lm} \right) G_{j_l h_l} \}.
\ee
Here $\tilde{C}$ denotes the average of $C$ over a spherical surface,  $a_{lm}$ and $b_{lm}$ are the so-called electric and magnetic scattering coefficients of an arbitrary sample while $C^{\sigma}_{lm}$ denotes the incident coefficients characterizing the nature of the wave. Moreover,
\begin{equation} \label{GPLQL}
G_{f_lg_l} = \frac{1}{2} \left[ 2f^*_l(u) g_l(u) + \frac{1}{u^2} \frac{\partial }{ \partial u} \left( u f^*_l(u)    \frac{\partial }{ \partial u}\left(ug_l(u)  \right) \right) \right].
\end{equation}
Here $G_{f_lg_l}$ is a scalar function that depends on spherical Bessel and Hankel functions. In particular, we may have $\{f_l, g_l \} =  \{j_l, j_l \}, \{j_l, h_l \}, \{h_l, h_l \}$, $j_l$ and $h_l$ being the spherical Bessel and Hankel functions, respetively~\cite{jackson1999electrodynamics}. Moreover, notice that $u = kr$ is the radius of the imaginary sphere, normalized by $\lambda/2\pi$, surrounding the object where the averaging integral is performed. For more details, please check Appendix~\ref{Orthogonality}.

\begin{figure}[t!]
    \centering
    \includegraphics[scale =0.575]{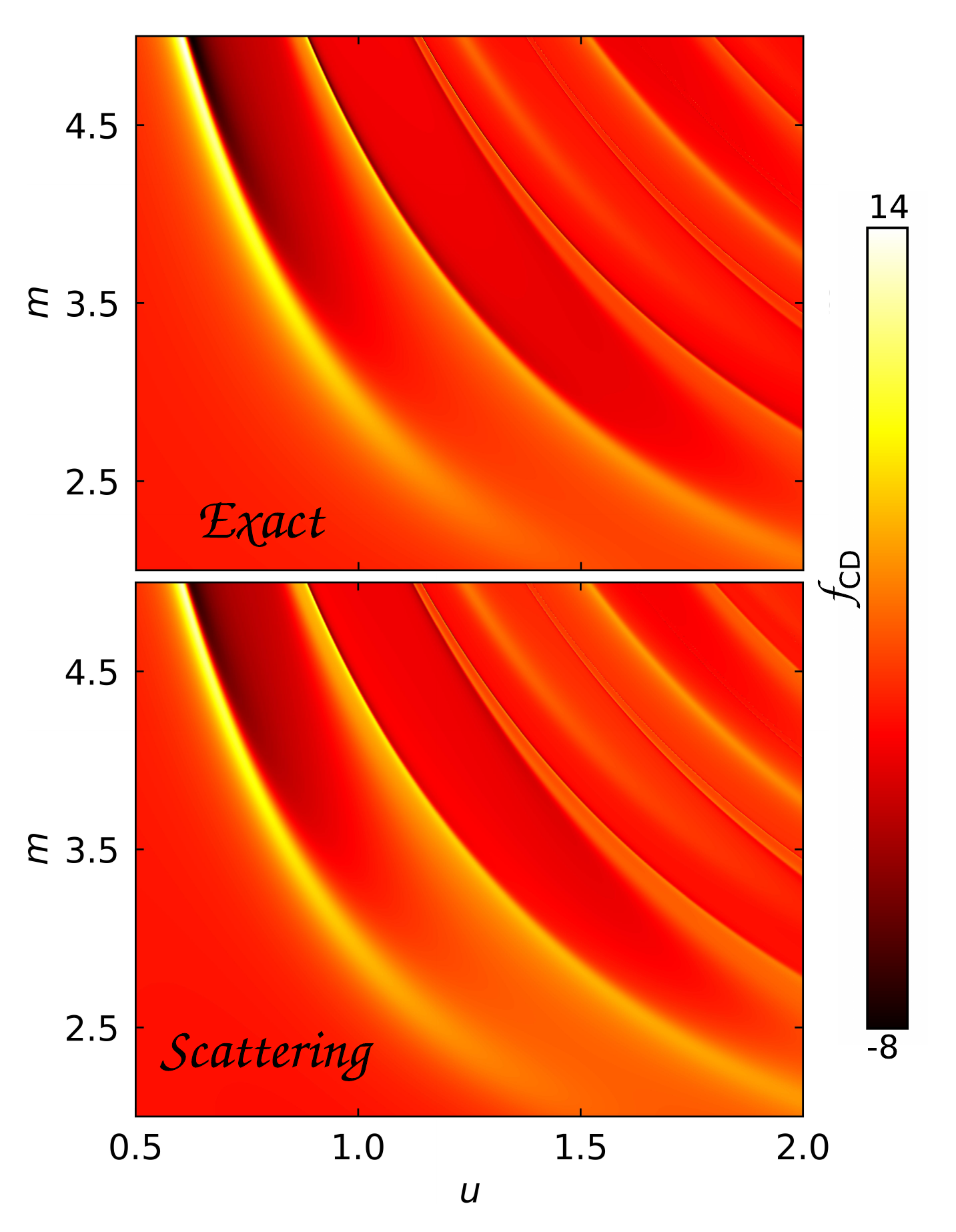}
    \caption{Circular dichroism enhancements for a multipolar sphere under the illumination of a circularly polarized plane-wave. a) $\mathit{f}_{\rm{CD}}$ (exact) and b) $\mathit{f}_{\rm{CD}} - {\tilde{\rm{C}}^{\sigma}_{\rm{int}}}/{\tilde{\rm{C}}^{\sigma}_{\rm{inc}}}$ (scattering). Here $u= x + \delta_x$, with $\delta_x \ll 1$, being $x=ka$ the optical size  and $m$ the index contrast.  }
    \label{fig_2}
\end{figure}

Equation~\eqref{f_CD_int_sep}, together with Eqs.~\eqref{C_incident}-\eqref{GPLQL}, is the first main result of this
paper. These equations describe the  integrated CD enhancement in the presence of scatterers of any size and shape under the excitation of fields with  well-defined helicity.
Our results overcome previous approximations, such as the widely used system of a circularly-polarized plane-wave illuminating dipolar objects~\cite{zhang2017amplification,yao2018enhancing,
hanifeh2020helicity, a2020chiral, hendry2012chiral, solomon2018enantiospecific,graf2019achiral, garcia2019enhanced, lasa2020surface, droulias2020absolute,mohammadi2021dual, rui2022surface}. %Note that previous works have restricted the analytical calculation of the surface-enhanced CD to dipolar spheres in the near-field limit (See Eq.~(9) of Ref.~\cite{garcia2013surface}). 
Thus, our findings may find applications in chiral sensing and chiral spectroscopy techniques beyond the current state-of-the-art.

At this point and for didactic purposes, we provide the steps to find $\mathit{f}_{\rm{CD}}$ around any optical antenna, which can be organized as follows: 
\begin{enumerate}
\item First, we need the solution of the electromagnetic fields under the illumination of a well-defined helicity beam. These can be obtained by any Maxwell's solver~\footnote{We notice that we can directly compute the $\mathit{f}_{\rm{CD}}$ factor for the exact solution of the electromagnetic fields under the illumination of a well-defined helicity beam. However, we would lose track of the role of the multipoles contributing to the $\mathit{f}_{\rm{CD}}$ factor, Moreover, imaginary integrating spheres surrounding the object would be needed at each calculation of the fields in order to perform the averaging integral surrounding the object.}.
\item  Then, we project in the far field the exact solution of the electromagnetic fields scattered by the object to obtain the scattering coefficients (see Eq. (9.123) in Jackson's book in its third edition~\cite{jackson1999electrodynamics}). For convenience, we  provide in Appendix~\ref{Conversion} the conversion between our scattering coefficients $a_{lm}$ and $b_{lm}$ to the ones employed in Jackson's book. Also notice that for spherical objects, we might use Mie's theory to directly jump to step 3.
\item Finally, we can  compute  $\mathit{f}_{\rm{CD}}$ via Eq.~\eqref{f_CD_int_sep}, together with Eqs.~\eqref{C_incident}-\eqref{GPLQL}.
\end{enumerate}

As an  illustrative example of the abovementioned recipe, we depict the exact expression of $\mathit{f}_{\rm{CD}}$ for a  sphere sustaining several multipoles under plane-wave illumination (see Fig.~\ref{fig_2}a)). Moreover, we also depict in Fig.~\ref{fig_2}b) the scattering contribution to $\mathit{f}_{\rm{CD}}$. That is, $\mathit{f}_{\rm{CD}} -{\tilde{\rm{C}}^{\sigma}_{\rm{int}}}/{\tilde{\rm{C}}^{\sigma}_{\rm{inc}}}$. In fact, we infer,  by comparing Fig.~\ref{fig_2}a) with Fig.~\ref{fig_2}b),  that the exact solution of $\mathit{f}_{\rm{CD}}$ can be fairly approximated to just scattering in near-field, namely, $\mathit{f}_{\rm{CD}} \sim 1 + {\tilde{\rm{C}}^{\sigma}_{\rm{sca}}}/{\tilde{\rm{C}}^{\sigma}_{\rm{inc}}}$. Now, we  understand the validity of this approximation based on the following facts:
\begin{itemize}
    \item On mathematical grounds, and according to Eq.~\eqref{Gplql}, it can be checked that $G_{h_l h_l} \gg |G_{j_l h_l}|$ for  $u < l$. That is, fundamental properties of spherical Bessel functions dictate that the scattering contribution  dominates over the interference term for $u<l$. 
    
      \item On physical grounds, we notice that there are no strong  resonances of the $l$-th multipole for $u>l$. %In other words, whenever $G_{h_l h_l} \sim |G_{j_l h_l}|$, the optical response of multipole $l$ does not dominate. 
      For example, dipolar and quadrupolar resonances typically arise for $u<1$ and $1<u<2$, respectively~\cite{coe2022unraveling}.
\end{itemize}

\begin{table*}[t!]
\begin{center}
\caption{Analytic expressions for the CD enhancement factor, $\mathit{f}_{\rm{CD}}$, depending on the interaction between an incident plane-wave and an achiral scatterer. Here $a_{lm}$ and $b_{lm}$  denote the electric and magnetic scattering coefficients and $\{G_{h_lh_l}, G_{j_lh_l}\}$ can be computed from Eq.~\eqref{Gplql};  $\langle\Lambda\rangle$ denotes the electromagnetic helicity expectation value and $\Lambda_\theta$ the helicity density; $\sigma_{\rm{sca}}$ and $\sigma_{\rm{ext}}$ are the scattering and extinction cross-sections; and $\lambda$ the radiation wavelength.}
\vspace{0.25cm}
\begin{tabular}{ c c c c}
\bf {Approximation in the calculation of $\mathit{f}_{\rm{CD}}$}
 & \bf {{Plane wave illumination}} \\
\hline
\\ 
{ Exact multipolar expansion} & $\mathit{f}_{\rm{CD}}  = 1 + \sum_{lm} \left(2l + 1 \right) \left( \text{Re} \{a_{lm} {b_{lm}}^* \} G_{h_l h_l} + \text{Re} \{  \left( a_{lm} + b_{lm} \right) G_{j_l h_l} \}\right)$ &
 \\ \\
 \hline
\\ 
{Scattering approximation} & $\mathit{f}_{\rm{CD}}  \sim 1 + \sum_{lm} \left(2l + 1 \right) \text{Re} \{a_{lm} {b_{lm}}^* \} G_{h_l h_l} $ &
 \\ \\
 \hline
\\ 
\makecell{Scattering approximation  for  arbitrary samples \\ well-described by a single multipolar order $l$} &  $\mathit{f}_{\rm{CD}}  \sim 1 + \dfrac{ \pi G_{h_l h_l}} {\lambda^2}    \langle \Lambda \rangle {\sigma}_{\rm{sca}} \underbrace{\longrightarrow}_{\text{Lossless}} \mathit{f}_{\rm{CD}}  \sim 1 + \dfrac{ \pi G_{h_l h_l}} {\lambda^2}  \langle \Lambda \rangle  {\sigma}_{\rm{ext}}$  \\ \\
 \hline
\\ 
\makecell{Scattering approximation for cylindrical samples \\ well-described by a single multipolar order $l$} & $\mathit{f}_{\rm{CD}}  \sim 1 + \dfrac{ \pi G_{h_l h_l}} {\lambda^2}    \Lambda_{\theta} {\sigma}_{\rm{sca}} \underbrace{\longrightarrow}_{\text{Lossless}} \mathit{f}_{\rm{CD}}  \sim 1 + \dfrac{ \pi G_{h_l h_l}} {\lambda^2}   \Lambda_{\theta}  {\sigma}_{\rm{ext}}$
 \\ \\
 \hline
\\ 
\end{tabular}
\label{table}
\end{center}
\end{table*}

%It is also important to notice that if we take into account that the $l$-electric-and-magnetic resonances appear for $u<l$, we can state that $\mathit{f}_{\rm{CD}} \sim 1 + { \tilde{\rm{C}}^{\sigma}_{\rm{sca}}}/{\tilde{\rm{C}}^{\sigma}_{\rm{inc}}}$ is, in general, a fair approximation.

At this point, we will delve into Eq.~\eqref{C_scattering}, which approximates the efficiency of an optical antenna to enhance the $\mathit{f}_{\rm{CD}}$ factor, as it has just been previously explained. In particular, let us examine its physical meaning when just one multipole order (electric and magnetic) contributes to the optical response of the antenna. In this regard, it is essential to note that the one multipole approximation includes the most studied scenario by the nanophotonic community devoted to enhanced chiral sensing: a circularly polarized plane wave incident on dipolar objects~\cite{zhang2017amplification,yao2018enhancing,
hanifeh2020helicity, a2020chiral, hendry2012chiral, solomon2018enantiospecific,graf2019achiral, garcia2019enhanced, lasa2020surface, droulias2020absolute,mohammadi2021dual, rui2022surface}. Now, when the scattering can be described by just one multipole order, $l$,  Eq.~\eqref{C_scattering} reads 
\be \label{C_sca_fixed}
\tilde{\rm{C}}^{\sigma}_{\rm{sca}}  &=&  \sigma G_{h_l h_l} \sum_{m=-l}^{m=l}|C^{\sigma }_{lm}|^2  \text{Re} \{a_{lm} {b_{lm}}^* \} ,
\ee 
By inspecting Eq.~\eqref{C_sca_fixed}, we notice that $\tilde{\rm{C}}^{\sigma}_{\rm{sca}}$ is proportional to the interference between the electric and magnetic scattering coefficients $\text{Re} \{a_{lm} {b_{lm}}^* \}$. Now, 
%let us introduce the $(\rm{V}/\rm{I})$-Stokes parameter,  which can be inferred  within polarization filters~\cite{crichton2000measurable}. That is,
let us introduce the normalized (and unit-less) electromagnetic helicity expectation value, which  reads~\cite{olmos2020kerker,olmos2020unveiling,lasa2022correlations}
\begin{equation} 
\langle \Lambda \rangle = \frac{\int_{\Omega} \left( {|\E^{\sigma+}_{\rm{sca}}}|^2 - {|\E^{\sigma -}_{\rm{sca}}}|^2\right)\; d \Omega}{\int_{\Omega} \left({|\E^{\sigma+}_{\rm{sca}}}|^2 + {|\E^{\sigma -}_{\rm{sca}}}|^2\right)\; d \Omega}.
\end{equation}
Computing $\langle \Lambda \rangle $ in the case in which the optical response can be described by a single multipolar order $l$~\cite{olmos2020kerker,olmos2020unveiling,lasa2022correlations}, we obtain
\begin{equation} \label{helicity}
{\langle \Lambda \rangle} =   2 \sigma  \frac{ \sum_{m} |C^{\sigma }_{lm}|^2 \text{Re} \{a_{lm} {b_{lm}}^* \}}{\sum_{m} |C^{\sigma }_{lm}|^2  \left(|a_{lm}|^2 +  |b_{lm}|^2\right)} =    \frac{ \sigma \sum_{m} |C^{\sigma }_{lm}| \text{Re} \{a_{lm} {b_{lm}}^* \}}{k^2 \sigma_{\rm{sca}}},
\end{equation}
where $\sigma_{\rm{sca}}$ is the scattering cross section~\cite{bohren2008absorption}.
At this point, we notice that the expression for $\tilde{\rm{C}}^{\sigma}_{\rm{sca}}$ resembles  $\langle \Lambda \rangle$. In fact and without loss of generality, we can write $\tilde{\rm{C}}^{\sigma}_{\rm{sca}}  = {G_{h_l h_l}} \langle \Lambda \rangle  k^2 \sigma_{\rm{sca}}$. As a result, $\mathit{f}_{\rm{CD}}$ yields
\be \label{Hel_sca}
\mathit{f}_{\rm{CD}} \sim 1 + \frac{{\tilde{\rm{C}}^{\sigma}_{\rm{sca}}}}{\tilde{\rm{C}}^{\sigma}_{\rm{inc}}} =  1 + \frac{{{G_{h_l h_l}} \langle \Lambda \rangle {k^2 \sigma_{\rm{sca}}}}}{\tilde{\rm{C}}^{\sigma}_{\rm{inc}}}.
\ee 
This is another significant result of the present work. We have linked the averaged optical chirality associated with scattered fields, $\tilde{\rm{C}}^{\sigma}_{\rm{sca}}$, which is usually computed in the near-field, with quantities that can be evaluated or measured in far-field, i.e. the helicity expectation value, $\langle \Lambda \rangle$ and the scattering cross-section, ${{\sigma}_{\rm{sca}}}$. 
In addition, and for helicity-preserving objects, $\langle \Lambda \rangle \sim 1$, the averaged optical chirality associated with scattered fields is simply given by the scattering cross-section. That is, $\tilde{\rm{C}}^{\sigma}_{\rm{sca}} \sim G_{h_l h_l} k^2 {\sigma}_{\rm{sca}}$.  It is also essential to notice that, for both lossless and helicity-preserving objects,  $\tilde{\rm{C}}^{\sigma}_{\rm{sca}} \sim  G_{h_l h_l} k^2{\sigma}_{\rm{ext}}$, with ${\sigma}_{\rm{ext}} = {\sigma}_{\rm{sca}}$, ${\sigma}_{\rm{ext}}$ being the extinction cross-section~\cite{bohren2008absorption}. That is,  the CD enhancement captured by the $\mathit{f}_{\rm{CD}}$ factor is related to a single measurement of the extincted power in the forward direction whenever $\langle \Lambda \rangle \sim 1$. The relation between $\mathit{f}_{\rm{CD}}$ and ${\sigma}_{\rm{ext}}$ for helicity-preserving achiral objects greatly reduces eventual experimental and numerical calculations devoted to enhanced chiral sensing.

So far, we have shown an alternative way to infer local CD enhancements in the near-field limit by computing far-field magnitudes such as the helicity expectation value, the scattering cross-section, or the extinction cross-section. In particular, a scenario of major interest for the purpose of enhanced chiral detection occurs when the antenna preserves the helicity of the incident illumination, i.e. whenever $\langle \Lambda \rangle \sim 1$. These objects satisfy $|\E^{\sigma \sigma' }_{\rm{sca}}(\r)| \sim 0$ for $\sigma \neq \sigma'$.
This phenomenon is desirable since the local sign of optical chirality is preserved.
Experimentally, identifying helicity-preserving scatterers requires measuring the polarization of all the scattered field components, something which is not feasible in practice. Thus, our question is: can we infer the helicity expectation value, $\langle \Lambda \rangle$, from a single measurement in the far-field? In the last part of this work, we will discuss scenarios in which the helicity density at a given scattering angle can be identical to its expected value. 
In particular, we will focus on cylindrically symmetric scatterers which preserve the total angular momentum in the incident direction ($m = m'$) and whose optical response can be well-described by a single multipolar order ($l = l'$), \textit{e.g.},  nanodisks at normal incidence or spherical particles under the illumination of tightly-focused Laguerre-Gaussian beams~\cite{sanz2021multiple}.

Mathematically, we can express the aforementioned condition as 
${\langle \Lambda \rangle} = \Lambda_{\theta}$, where $\Lambda_{\theta}$
denotes the helicity density at an angle $\theta$, where $\theta$ is the scattering angle. After some algebra (see Appendix~\ref{Helicity_same}), we obtain:
\begin{equation} \label{Todo_angle}
P^{m}_{l} (\cos \theta) \frac{\partial P^{m}_{l} (\cos \theta)}{\partial \cos \theta} = 0~~~\Longrightarrow~~~{\langle \Lambda \rangle}  = \Lambda_{\theta},
\end{equation}
where $P^{m}_l(\cos \theta)$ are the associated Legendre Polynomials~\cite{jackson1999electrodynamics}.
This is another key result of the present work. The helicity expectation value can indeed be computed from a single measurement of the helicity density at a specific angle $\theta$. Our result implies that for a cylindrically symmetric scatterer whose response can be well-described by a single multipolar order, $l$, if we excite it with an illumination with a fixed total angular momentum, $m$, Eq. \eqref{Todo_angle} specifies the angle at which the helicity density is equal to its expected value.
For instance, if we consider the typical case of a cylindrically symmetric dipolar target ($l = 1$) under a circularly polarized plane-wave illumination ($m = \pm 1$), Eq.~\eqref{Todo_angle} yields that the angle we should look at is $\theta = \pi/2$. That is, the expectation value of the electromagnetic helicity can be inferred from a single measurement at the right angle. Thus, and according to Eq.~\eqref{Hel_sca}, for lossless and cylindrically symmetric targets, we can infer the $\mathit{f}_{\rm{CD}}$ factor by just considering two far-field measurements: extinction cross-section, in the forward direction, and helicity density, at an angle $\theta$ specified by Eq. \eqref{Todo_angle}~\footnote{In terms of the usual Stokes parameters, extinction is proportional to the $s_0$ parameter and helicity density is the ratio $s_3/s_0$~\cite{lasa2022correlations}.}.

Table~\ref{table} resumes the main results of this work, particularized for plane-wave illumination, as it is the most common external excitation for enhanced chiral sensing to date. In short, we have derived an exact multipole expansion of the integrated CD enhancement factor, $\mathit{f}_{\rm{CD}}$, for scatterers of any form and shape under general illumination conditions. In addition, we have established a roadmap to infer local CD enhancements from far-field measurements. That is, $\mathit{f}_{\rm{CD}}$ can be extracted by calculating the helicity expectation value and the scattering cross section, two Stokes parameters that can be evaluated in the far-field limit. Finally, we have shown an even more practical route to deduce $\mathit{f}_{\rm{CD}}$ for cylindrically symmetric objects by means of measuring the extinction cross-section and the local density of electromagnetic helicity at specific angles. Our results pave the way for experimental verification and characterization of building blocks for CD enhancement from far-field measurements, and thus, may give rise to novel developments in the field of chiral light-matter interactions.

\bibliography{Bib_tesis}
\clearpage

\onecolumngrid
\appendix

\section{Multipolar electromagnetic fields in a well-defined helicity basis} \label{Fields}

\subsection{Incident electromagnetic fields}
\label{incident_Fields}
To start with the calculation of the surface-enhanced circular dichroism (CD),  we need first to write down the  electromagnetic fields. The most generic expression of the incident electromagnetic fields are given by
\be \label{E_inc}
\E_{\rm{inc}}( \r) &=&  \sum_{l=0}^\infty \sum_{m=-l}^{+l}  g^{e}_{lm}   \bm{N}^{j}_{lm}( \r) + g^{m}_{lm}   \bm{M}^{j}_{lm}( \r), \\
 \im Z\H_{\rm{inc}}( \r) &=&  \sum_{l=0}^\infty \sum_{m=-l}^{+l}  g^{e}_{lm}   \bm{M}^{j}_{lm}( \r) + g^{m}_{lm}   \bm{N}^{j}_{lm}( \r),
\ee
where $g^{e}_{lm}$ and $g^{m}_{lm}$ stands for the incident electric and magnetic beam's shape coefficients, respectively, and 
\be \label{Multipoles_j}
{\boldsymbol{M}}^{j}_{lm}(\r)  = j_l(kr)\boldsymbol{X}_{lm}(\r),  \qquad
{\boldsymbol{N}}^{j}_{lm}(\r)  = \frac{\boldsymbol{\nabla} \times {\boldsymbol{M}}^{j}_{lm}(\r) }{k},  \qquad
\boldsymbol{X}_{lm}(\r) = \frac{{\bf{L}}Y_{lm} (\theta,\varphi)}{\sqrt{l(l+1)}}.
\ee
Here $\boldsymbol{M}^{j}_{lm}(\r) $ and $\boldsymbol{N}^{j}_{lm}(\r) $  are (incident) Hansen's multipoles~\cite{jackson1999electrodynamics}, $ j_l(kr)$ are the spherical Bessel functions, $k$ being the radiation wavelength, and $\r$ the observation point. Moreover, 
$Y_{lm} (\theta,\varphi)$ are the spherical harmonics, $\theta$ and $\varphi$ being  the polar and azimuthal angles,  and $ {\bf{L}} =  \left\{ -\im \r \times \boldsymbol{\nabla}\right\} $ is the total angular momentum operator. At this point, let us consider an arbitrary incident electromagnetic field with well-defined helicity, $\sigma = \pm 1$. Mathematically, we can write this well-defined helicity field as
\be \label{E_inc_Silberstein}
\E^{\sigma}_{\rm{inc}}( \r) &=& \frac{\E_{\rm{inc}}( \r) + \sigma \im Z\H_{\rm{inc}}( \r)}{2} = \sum_{l=0}^\infty \sum_{m=-l}^{+l}  \left(\frac{ {g^{e}_{lm} + \sigma  g^{m}_{lm}}}{\sqrt{2}} \right) \left(  \frac{ {\boldsymbol{N}}^{j}_{lm} (\r) +  \sigma {\boldsymbol{M}}^{j}_{lm}(\r)  }{\sqrt{2}}  \right)   = \sum_{l=0}^\infty \sum_{m=-l}^{+l} C_{lm}^{ \sigma} \boldsymbol{\Psi}_{lm}^{\sigma} ( \r),
\ee
where 
\begin{align} \label{C_lmsigmas}
 C_{lm}^{ \sigma} = \frac{ {g^{e}_{lm} + \sigma  g^{m}_{lm}}}{\sqrt{2}}, && \text{and} && \boldsymbol{\Psi}_{lm}^{\sigma} ( \r) =  \frac{ {\boldsymbol{N}}^{j}_{lm} (\r) +  \sigma {\boldsymbol{M}}^{j}_{lm}(\r)  }{\sqrt{2}}.
\end{align}
Let us recall that the multipoles $ \boldsymbol{\Psi}_{lm}^{\sigma}(\r) $ 
are eigenvectors of the squared angular momentum  $L^2$, the projection of the angular momentum on one direction, $L_z$, and helicity $\bm{\Lambda} = (1/k) \nabla \times$~\cite{fernandez2013electromagnetic} with eigenvalues $l(l+1)$, $m$, $\sigma$, respectively.

\subsection{Scattered and total electromagnetic fields}
\label{Scattered_Fields}

At this point, let us consider the scattered electromagnetic fields. The most generic expression of these fields is given by
\be \label{E_sca}
\E^{\sigma}_{\rm{sca}}( \r) &=&  \sum_{l=0}^\infty \sum_{m=-l}^{+l}  a^{\sigma}_{lm}   \bm{N}^{h}_{lm}( \r) + b^{\sigma}_{lm}   \bm{M}^{h}_{lm}( \r), \\
 \im Z\H^\sigma_{\rm{sca}} ( \r)&=&  \sum_{l=0}^\infty \sum_{m=-l}^{+l}  a^{\sigma}_{lm}   \bm{M}^{h}_{lm}( \r) + b^{\sigma}_{lm}   \bm{N}^{h}_{lm}( \r),
\ee
where $a^{\sigma}_{lm}$ and $b^{\sigma}_{lm}$ stand for the  electric and magnetic scattering coefficients, respectively. Notice that $a^{\sigma}_{lm}$ and $b^{\sigma}_{lm}$ depend on the incident illumination. As a result, we have explicitly indicated the  $\sigma$-dependency.
Moreover,  
\be \label{Multipoles_h}
{\boldsymbol{M}}^{h}_{lm}(\r)  = h_l(kr)\boldsymbol{X}_{lm}(\r),  \qquad
{\boldsymbol{N}}^{h}_{lm}(\r)  = \frac{\boldsymbol{\nabla} \times {\boldsymbol{M}}^{h}_{lm}(\r) }{k},
\ee
where $\boldsymbol{M}^{h}_{lm}(\r) $ and $\boldsymbol{N}^{h}_{lm}(\r) $  are (scattered) Hansen's multipoles~\cite{jackson1999electrodynamics} and $ h_l(kr)$ are the spherical Hankel functions. Following the steps done in Eq.~\eqref{E_inc_Silberstein}, the electric field reads as
\be \label{E_sca_Silberstein}
\E^{\sigma \sigma'}_{\rm{sca}}( \r) = \frac{\E^\sigma_{\rm{sca}}( \r) + \sigma' \im Z\H^\sigma_{\rm{sca}}( \r)}{2} &=& \sum_{l=0}^\infty \sum_{m=-l}^{+l}  \left(\frac{ {a^{\sigma}_{lm} + \sigma' b^{\sigma}_{lm}}}{\sqrt{2}} \right) \left(  \frac{ {\boldsymbol{N}}^{h}_{lm} (\r) +  \sigma' {\boldsymbol{M}}^{h}_{lm}(\r)  }{\sqrt{2}}  \right)   = \sum_{l=0}^\infty \sum_{m=-l}^{+l} D_{lm}^{\sigma  \sigma'} \boldsymbol{\Phi}_{lm}^{\sigma'} ( \r),
\ee
where
\begin{align} \label{D_lmsigmas}
D_{lm}^{\sigma  \sigma'} = \frac{ {a^{\sigma}_{lm} + \sigma' b^{\sigma}_{lm}}}{\sqrt{2}}, && \text{and} && \boldsymbol{\Phi}_{lm}^{\sigma'} ( \r) =  \frac{ {\boldsymbol{N}}^{h}_{lm} (\r) +  \sigma' {\boldsymbol{M}}^{h}_{lm}(\r)  }{\sqrt{2}}.
\end{align}
%The scattered electromagnetic fields read then as, \be 
%\E^{\sigma}_{\rm{sca}}( \r) &=& \sum_{\sigma'=\pm1} \E^{ \sigma \sigma'}_{\rm{sca}}( \r), \qquad \im Z\H^{\sigma}_{\rm{sca}}( \r) = \sum_{\sigma'=\pm1} \sigma' \E^{ \sigma \sigma'}_{\rm{sca}}( \r).
%\ee 
At this stage, let us write down the relation between the incident and scattering amplitudes. These are given by Cramer's rule of the tangential Maxwell boundary conditions~\cite{jackson1999electrodynamics}, $
a^{\sigma}_{lm} =  a_{lm} g^{e}_{lm}$ and  $b^{\sigma}_{lm} =  b_{lm} g^{m}_{lm}$
where $a_{lm}$ and  $b_{lm}$ are the so-called electric and magnetic scattering coefficients, respectively. Notice $a_{lm}$ and  $b_{lm}$  do not depend on the incident illumination but on the optical properties and geometry of the target, \textit{e.g.}, the electric and magnetic Mie coefficients.  
Now, and
since we are dealing with a well-defined helicity field, we can notice from Eq.~\eqref{E_inc_Silberstein} that $g^{e}_{lm} = \sigma  g^{m}_{lm}$. Accordingly,  we can write 
\begin{align} \label{Amplitudes}
a^{\sigma}_{lm} =  a_{lm} \left( \frac{C^{\sigma}_{lm}}{\sqrt{2}} \right), && b^{\sigma}_{lm} =  \sigma b_{lm} \left( \frac{C^{\sigma}_{lm}}{\sqrt{2}} \right).
\end{align}
Now, by inserting Eq.~\eqref{Amplitudes} into the left side of Eq.~\eqref{D_lmsigmas}, we arrive to
\be \label{Dlm}
D^{\sigma \sigma '}_{lm} =C^{\sigma}_{lm} \frac{ \left( {a_{lm} + \sigma \sigma' b_{lm}}\right)}{2}.
\ee
From now on, this will be our choice for the representation of the scattered coefficients. 
To conclude Appendix~\ref{Fields}, let us write the total electromagnetic fields. These are given by the additive sum of the incident and scattered electromagnetic fields. Hence, by taking into account both Eq.~\eqref{E_inc_Silberstein} and Eq.~\eqref{E_sca_Silberstein}, we can straightforwardly write,
\be \label{E_total_Silberstein}
\E^{\sigma}_{\rm{tot}}( \r) &=& \sum_{\sigma'=\pm1} \E^{ \sigma \sigma'}_{\rm{tot}}( \r), \qquad \im Z\H^{\sigma}_{\rm{tot}}( \r) = \sum_{\sigma'=\pm1} \sigma' \E^{ \sigma \sigma'}_{\rm{tot}}( \r),
\ee 
with
\be\label{E_total}
\E^{\sigma \sigma'}_{\rm{tot}}(\r) &=&  \E^{\sigma \sigma'}_{\rm{sca}}(\r) + \E^{\sigma}_{\rm{inc}}(\r) \delta_{\sigma \sigma'}, 
\ee 
where $\delta_{\sigma \sigma'}$ is a Kronecker delta.
Next, we will use the orthogonality expressions that satisfy both the incident and scattered electromagnetic fields to compute the exact multipolar expansion of the CD enhancement factor.

\section{An exact multipolar expansion of $f_{\rm{CD}}$ beyond the plane-wave picture} 
\subsection{An exact multipolar expansion of $f_{\rm{CD}}$: Orthogonality relations of well-defined helicity multipoles}
\label{Orthogonality}
To derive an exact multipolar expansion of the CD enhancement factor, $\mathit{f}_{\rm{CD}}$, beyond the plane-wave picture, we need to know the  orthogonality relations that satisfy both incident and scattered electromagnetic fields over an integrating sphere surrounding the object under illumination.  To that end, we need first to calculate the set of orthogonality relations that fulfill the multipoles with well-defined helicity. That is, we need the orthogonality relations between incident $\{ \boldsymbol{\Psi}_{lm}^{\sigma}, \boldsymbol{\Psi}_{l'm'}^{\sigma} \}$, interference  $\{ \boldsymbol{\Psi}_{lm}^{\sigma}, \boldsymbol{\Phi}_{l'm'}^{\sigma} \}$, and scattering $\{ \boldsymbol{\Phi}_{lm}^{\sigma}, \boldsymbol{\Phi}_{l'm'}^{\sigma}\}$ terms, according to Eq.~\eqref{f_CD_int}. Fortunately, all these relations can be derived from Jackson's third edition book. Let us start this section by transcribing Eq.~$10.48$ that can be found on page $472$ of Ref.~\cite{jackson1999electrodynamics}:
\begin{align} \label{Jackson_ortho}
\int_{\Omega} {N^{f}_{lm}}^* \cdot N^{g}_{l'm'} d \Omega =  \delta_{ll'} \delta_{mm'} \left( f^*_l(u) g_l(u) +  \frac{1}{u^2} \frac{\partial }{ \partial u} \left( u f^*_l(u)    \frac{\partial }{ \partial u}\left(ug_l(u)  \right) \right) \right), && \int_{\Omega} {M^{f}_{lm}}^* \cdot M^{g}_{l'm'} d \Omega = f^*_l(u) g_l(u)   \delta_{ll'} \delta_{mm'}.
\end{align}
Here  $u = kr$ denotes the optical radius of the integrating sphere and $\{ \delta_{ll'}, \delta_{mm'}\}$ are Kronecker deltas. Notice that $\{ f_l(u), g_l(u)\}$ denote either Bessel or Hankel spherical  functions, namely, $j_l(u)$ and $h_l(u)$~\cite{jackson1999electrodynamics}, depending on the nature of the field: incident or scattered, respectively. 

At this point, we have already learned that multipoles with well-defined helicity $\{ \boldsymbol{\Psi}_{lm}^{\sigma}, \boldsymbol{\Phi}_{lm}^{\sigma} \}$ are constructed by a linear combination of the Hansel multipoles (see the right side of Eq.~\eqref{C_lmsigmas} and Eq.~\eqref{D_lmsigmas} . As a result,  we can write from Eq.~\eqref{Jackson_ortho}
\begin{align} \label{Orthogonality relations}
 \int_{\Omega}({\boldsymbol{\Psi}^{\sigma}_{l'm'}})^* \cdot \boldsymbol{\Psi}_{lm}^{\sigma} \;
  d \Omega =   G_{j_lj_l} \delta_{ll'}\delta_{mm'}, &&
 \int_{\Omega} ({\boldsymbol{\Phi}^{\sigma}_{l'm'}})^* \cdot {\boldsymbol{\Phi}^{\sigma}_{lm}} 
 \; d \Omega =   G_{h_lh_l} \delta_{ll'}\delta_{mm'}, &&
 \int_{\Omega} ({\boldsymbol{\Psi}^{\sigma}_{l'm'}})^* \cdot \boldsymbol{\Phi}_{lm}^{\sigma}  \; d \Omega =   G_{j_lh_l} \delta_{ll'}\delta_{mm'},
\end{align}
with
\begin{equation} \label{Gplql}
G_{f_lg_l} = \frac{1}{2} \left[ 2f^*_l(u) g_l(u) + \frac{1}{u^2} \frac{\partial }{ \partial u} \left( u f^*_l(u)    \frac{\partial }{ \partial u}\left(ug_l(u)  \right) \right) \right].
\end{equation}
Now,  we can re-write Eq.~\eqref{Gplql} to get rid of second derivatives by making use of fundamental properties of the Ricatti-Bessel functions~\cite{watson1995treatise}. In particular, we can write,  
\begin{align}
G_{j_lj_l} = \frac{1}{2} \left[ j^2_l(u) \left(1 + \frac{l(l+1)}{u^2} \right) + \frac{1}{u^2} \left( \frac{\partial }{ \partial u}\left(uj_l(u)   \right), \right)^2 \right] &&
G_{h_lh_l} = \frac{1}{2} \left[ |h_l(u)|^2 \left(1 + \frac{l(l+1)}{u^2} \right) + \frac{1}{u^2} \left| \frac{\partial }{ \partial u}\left(uh_l(u)   \right) \right|^2 \right],
\end{align}

\be
G_{j_lh _l} = \frac{1}{2} \left[ j_l(u) h_l(u)\left(1 + \frac{l(l+1)}{u^2} \right) + \frac{1}{u^2} \left( \frac{\partial }{ \partial u}\left(uj_l(u)   \right) \frac{\partial }{ \partial u}\left(uh_l(u)   \right) \right) \right],
\ee
where $\frac{\partial }{ \partial u}\left(uf_l(u)   \right) ) = uf_{l-1}(u) - lf_l(u)$ is satisfied for $f(u) = \{j_l(u), h_l(u) \}$.

\subsection{An exact multipolar expansion of $f_{\rm{CD}}$: From the helicity basis to the standard electric and magnetic multipolar expansion} \label{EXACT}

At this stage, we have all ingredients to calculate the exact multipolar expansion of  $f_{\rm{CD}}$.  The starting point of this section will be Eq.~\eqref{f_CD_int}. According  to Eq.~\eqref{f_CD_int},  we need the orthogonality relations that satisfy the Rieman-Silberstein representation of the incident and scattered electromagnetic fields:
\be
\tilde{\rm{C}}^{\sigma}_{\rm{inc}} &=& \int_{\Omega}
{\rm{C}}^{\sigma}_{\rm{inc}}(\r)  d \Omega = \int_{\Omega}
\sigma |\E^{\sigma}_{\rm{inc}}(\r)|^2 d \Omega, \\ 
\tilde{\rm{C}}^{\sigma}_{\rm{sca}} &=&
\int_{\Omega} 
{\rm{C}}^{\sigma}_{\rm{sca}}(\r)  d \Omega = \int_{\Omega} 
|\E^{\sigma + }_{\rm{sca}}(\r)|^2 - |\E^{\sigma - }_{\rm{sca}}(\r)|^2 d \Omega, \\
{\rm{C}}^{\sigma}_{\rm{int}} &=&
 \int_{\Omega}
\tilde{{\rm{C}}}^{\sigma}_{\rm{int}}(\r)  d \Omega =  2 \sigma \int_{\Omega}
 \text{Re} \{{\E^{\sigma}_{\rm{inc}}}^* ( \r ) \cdot {\E^{\sigma \sigma}_{\rm{sca}}} ( \r ) \}  d \Omega.
\ee
These orthogonality relations can be computed by combining  Eq.~\eqref{E_inc_Silberstein} and Eq.~\eqref{E_sca_Silberstein}  with Eq.~\eqref{Orthogonality relations}. In fact and after some algebraic manipulations, it can be shown that 
\begin{align}
\tilde{\rm{C}}^{\sigma}_{\rm{inc}} =  \sigma  \sum_{lm}
|C^{\sigma }_{lm}|^2 G_{j_l j_l}, &&
\label{C_sca_sca}
\tilde{\rm{C}}^{\sigma}_{\rm{sca}} = \sum_{lm } \left( |D^{\sigma +}_{lm}|^2 - |D^{\sigma -}_{lm}|^2 \right) G_{h_l h_l}, &&
\tilde{{\rm{C}}}^{\sigma}_{\rm{int}} = 2 \sigma \sum_{lm}  \text{Re} \{{C^\sigma_{lm}}^* D^{\sigma \sigma}_{lm} G_{j_l h_l} \}.
\end{align}
Now, let us insert Eq.~\eqref{Dlm} into Eq.~\eqref{C_sca_sca} to obtain a closed-relation of the optical chirality enhancements in terms of the electric and magnetic scattering coefficients. After some algebra, we arrive to 
\begin{equation} 
\mathit{f}_{\rm{CD}} = 1 + \frac{ \tilde{\rm{C}}^{\sigma}_{\rm{sca}} + \tilde{\rm{C}}^{\sigma}_{\rm{int}}}{\tilde{\rm{C}}^{\sigma}_{\rm{inc}}},
\end{equation}
with
\begin{align}
\tilde{\rm{C}}^{\sigma}_{\rm{inc}}  = \sigma  \sum_{lm}
|C^{\sigma }_{lm}|^2 G_{j_l j_l}, &&
\tilde{\rm{C}}^{\sigma}_{\rm{sca}}  = \sigma \sum_{lm} |C^{\sigma }_{lm}|^2  \text{Re} \{a_{lm} {b_{lm}}^* \} G_{h_l h_l}, && 
\tilde{\rm{C}}^{\sigma}_{\rm{int}} =   \sigma \sum_{lm} |C^{\sigma }_{lm}|^2  \text{Re} \{  \left( a_{lm} + b_{lm} \right) G_{j_l h_l} \}.
\end{align}

\section{Conversion from our
conventions to those in Jackson’s book} \label{Conversion}
The multipolar expansion of the scattered electromagnetic fields provided in Jackson's book reads as~\cite{jackson1999electrodynamics} 
\be \label{Jack_e}
\E^{\rm{Jack}}_{\rm{sca}} &=&  Z  \sum_{l=0}^\infty \sum_{m=-l}^{+l}  \im a_{E}(l, m)   \bm{N}^{h}_{lm}( \r) + a_{M}(l, m)   \bm{M}^{h}_{lm}( \r), \\
\label{Jack_m}
 \im Z\H^{\rm{Jack}}_{\rm{sca}} &=&   Z  \sum_{l=0}^\infty \sum_{m=-l}^{+l}  \im a_{E}(l, m)   \bm{M}^{h}_{lm}( \r) + a_{M}(l, m)   \bm{N}^{h}_{lm}( \r).
\ee
Now, by inspecting Eqs.~\eqref{Jack_e}-\eqref{Jack_m}, we notice that the conversion from our
conventions to those in Jackson’s book are given by 
\begin{align}
a_{lm} = \sqrt{2} \im  Z  \left[\frac{a_{E}(l, m)}{C^{\sigma}_{lm}} \right], && b_{lm} = \sqrt{2} \sigma Z  \left[\frac{a_{M}(l, m)}{C^{\sigma}_{lm}} \right].
\end{align}
where $Z = \mu/ \epsilon$ is the medium impedance.
Notice that the electric and magnetic coefficients, provided in Jackson's book, can be computed by conventional  far-field projections of the scattered electromagnetic fields (see Eq. (9.123) in Ref.~\cite{jackson1999electrodynamics}). For completeness, we transcribe these expressions,
\begin{align}
Z a_{E}(l, m) h_l(kr) =- \frac{k}{\sqrt{l(l+1)}} \int Y^*_{lm}(\theta, \varphi) \r \cdot \E^{\rm{Jack}}_{\rm{sca}}, && a_{M}(l, m) h_l(kr) = \frac{k}{\sqrt{l(l+1)}} \int Y^*_{lm}(\theta, \varphi) \r \cdot \H^{\rm{Jack}}_{\rm{sca}}.
\end{align}

\section{Extracting the helicity expectation value from a single measurement of its local density}
\label{Helicity_same}
In this Appendix, we derive the condition given in Eq. \eqref{Todo_angle}, that relates the local density of helicity at a certain angle, $\Lambda_{\theta, \phi}$, with the helicity expectation value, $\langle\Lambda\rangle$. For that aim, we should first define the local density of helicity, which we consider to be given in far-field by:
\begin{equation}
    \label{Ldensity}
    \Lambda_{\theta, \varphi} \equiv \lim_{kr \rightarrow \infty} \frac{|\mathbf{E}^{\sigma+}_{\rm{sca}}(r, \theta, \varphi)|^2 - |\mathbf{E}^{\sigma-}_{\rm{sca}}(r, \theta, \varphi)|^2}{|\mathbf{E}^{\sigma+}_{\rm{sca}}(r, \theta, \varphi)|^2 + |\mathbf{E}^{\sigma-}_{\rm{sca}}(r, \theta, \varphi)|^2},
\end{equation}
where $\mathbf{E}^{\sigma\sigma'}_{\rm{sca}}(r, \theta, \varphi)$ is the  scattered field written in terms of electromagnetic modes with well-defined helicity. From Eq.~\eqref{Ldensity}, we  notice that we require the asymptotic behavior of $\mathbf{E}^{\sigma\sigma'}_{\rm{sca}}(r, \theta, \varphi)$  in the far-field limit. For that aim, we need first to calculate how Hansel multipoles behave in the far-field limit. After some algebra, we arrive to
\begin{align}
\lim_{kr \rightarrow \infty} \bm{N}^{h}_{lm}( \r) = -\frac{e^{ikr}}{kr}   \frac{(-i)^{l+1}}{\sqrt{l(l+1}} \bm{\xi}_{lm}(\theta, \varphi), &&
\lim_{kr \rightarrow \infty} \bm{M}^{h}_{lm}( \r) = -\frac{e^{ikr}}{kr}   \frac{(-i)^{l+1}}{\sqrt{l(l+1}} i \bm{\eta}_{lm}(\theta, \varphi),
\end{align}
where $\bm{\xi}_{lm}(\theta, \varphi) = r\nabla Y_{lm}(\theta, \varphi)$ and $\bm{\eta}_{lm}(\theta, \varphi) = \hat{r} \times  \bm{\xi}_{lm}(\theta, \varphi)$. Now, the far-field expression of the electromagnetic field scattered by an arbitrary sample can be computed \cite{Carrascal}:
\begin{equation}
    \label{Fsca}
    \mathbf{E}^{\sigma\sigma'}_{\rm{sca}}(r, \theta, \varphi) = -\frac{e^{ikr}}{kr}\sum_{lm}(-i)^{l+1}\frac{C_{lm}^\sigma}{\sqrt{2l(l+1)}}\left[\frac{a_{lm} + \sigma\sigma'b_{lm}}{\sqrt{2}}\right]\left[\frac{\bm{\xi}_{lm} + i\sigma'\bm{\eta}_{lm}}{\sqrt{2}}\right].
\end{equation}
%where $\boldsymbol{\Psi}_{lm}(\theta, \varphi) = r\nabla Y_{lm}(\theta, \varphi)$ and $\boldsymbol{\Phi}_{lm}(\theta, \varphi) = \hat{r}\times\boldsymbol{\Psi}_{lm}(\theta, \varphi)$, with $Y_{lm}(\theta, \varphi)$ the scalar spherical harmonics.
Substituting the scattered field in Eq. \eqref{Fsca} into the expression of the helicity density given by Eq. \eqref{Ldensity}, we obtain for fixed $l$ and $m$ values:
\begin{equation}
    \label{L1}
    \Lambda_\theta = 2\sigma \frac{\text{Re}\left(a_{lm}{b_{lm}}^*\right)\left[|\bm{\xi}_{lm}|^2 + |\bm{\eta}_{lm}|^2\right] - \left[|a_{lm}|^2 + |b_{lm}|^2\right]\text{Im}\left( \bm{\xi}_{lm}^* \cdot \bm{\eta}_{lm} \right)}{\left[|a_{lm}|^2 + |b_{lm}|^2\right]\left[|\bm{\xi}_{lm}|^2 + |\bm{\eta}_{lm}|^2\right] - 4\text{Re}\left(a_{lm}{b_{lm}}^*\right)\text{Im}\left( \bm{\xi}_{lm}^* \cdot \bm{\eta}_{lm} \right)}.
\end{equation}
The type of scatterers which may be described by fixed $l$ and $m$ values are cylindrically symmetric particles, illuminated by a beam with a well-defined angular momentum, $m$, and with a non multipolar response. Due to the cylindrical symmetry of the scatterers, helicity density cannot depend on $\varphi$ variable. This is the reason why we have chosen to write helicity density as $\Lambda_\theta$ in Eq. \eqref{L1}. Crucially, it can be checked that whenever $\text{Im}\left( \bm{\xi}_{lm}^* \cdot \bm{\eta}_{lm} \right) = 0$, one recovers the expression of the helicity expectation value, i.e.
\begin{equation}
    \text{Im}\left( \bm{\xi}_{lm}^* \cdot \bm{\eta}_{lm} \right) = 0~~~\Longrightarrow~~~\Lambda_\theta = \langle \Lambda \rangle = 2\sigma \frac{\text{Re}\left(a_{lm}{b_{lm}}^*\right)}{|a_{lm}|^2 + |b_{lm}|^2}.
\end{equation}
Importantly, the condition $\text{Im}\left( \bm{\xi}_{lm}^* \cdot \bm{\eta}_{lm} \right) = 0$ is purely geometrical, i.e. does not depend on the particular response of the scatterer. This fact makes the expression completely general and applicable to any type of cylindrical sample whose response is well-described by a fixed $l$. Thus, for this type of scatterers, there are locations in the far-field at which the helicity density is equal to the helicity expectation value.

The specific sites at which $\langle\Lambda\rangle = \Lambda_\theta$ are obtained by finding the solutions to the equation $\text{Im}\left( \bm{\xi}_{lm}^* \cdot \bm{\eta}_{lm}\right) = 0$. More explicitly, we have that the vector and scalar spherical harmonics are written as:
\begin{align}
    \bm{\xi}_{lm}(\theta,\varphi) &= \frac{\partial Y_{lm}(\theta, \varphi)}{\partial\theta}\hat{u}_\theta + \frac{1}{\sin \theta}\frac{\partial Y_{lm}(\theta,\varphi)}{\partial\varphi}\hat{u}_\varphi\\
    \bm{\eta}_{lm}(\theta,\varphi) &= \frac{\partial Y_{lm}(\theta,\varphi)}{\partial\theta}\hat{u}_\varphi - \frac{1}{\sin\theta}\frac{\partial Y_{lm}(\theta, \varphi)}{\partial\varphi}\hat{u}_\theta\\
    Y_{lm}(\theta, \varphi) &= \sqrt{\frac{2l + 1}{4\pi}\frac{(l-m)!}{(l+m)!}}P_l^m(\cos\theta)e^{im\varphi},
\end{align}
where $P_l^m(\cos \theta)$ are the associated Legendre polynomials. With the definitions above it is straightforward to check that
\begin{equation}
    P_l^m(\cos\theta)\frac{\partial P_l^m(\cos\theta)}{\partial \cos\theta} = 0~~~\Longrightarrow~~~\text{Im}\left( \bm{\xi}_{lm}^* \cdot \bm{\eta}_{lm} \right) = 0.
\end{equation}
In conclusion, for fixed values of $l$ and $m$, there is an angle $\theta$, given by the transcendental equation above, at which the helicity density, $\Lambda_\theta$, is equal to the helicity expectation value, $\langle\Lambda\rangle$.

%\section{Near-field limit of $G_{h_l h_l}$} 
%\label{Aitzol}
%Let us consider the following limit,
%\be 
%\lim_{u \ll 1} G_{h_lh_l} = \lim_{u \ll 1} \frac{1}2 \left[ 2|h_l(u)|^2 +  \frac{1}{ u^2} \frac{\partial }{ \partial u} \left( u h^*_l(u)    \frac{\partial }{ \partial u}\left(uh_l(u)  \right) \right) \right] = \lim_{u \ll 1} \frac{ \left[ |h_l(u)|^2 + l \left(l + 1 \right)|h_l(u)|^2 \right]}{2u^2},
%\ee
%where $u = kr$.
%Now, let us consider $l =1$ in the previous expression. That is 
%\be 
%\lim_{kr \ll 1} G_{h_1h_1} = \lim_{kr \ll 1} \frac{3}{2} %\left( \frac{|h_1(kr)|^2}{(kr)^2} \right)= \frac{3}{2} %\left( \frac{1}{ (kr)^6} \right), \qquad \text{since} %\qquad h_1 (u) = \frac{e^{-\im u}}{u} \left(1 + \frac{1}%{u} \right).
%\ee

\end{document}